\documentclass[12pt]{article}
\usepackage{amsmath,amsfonts,amssymb}
\usepackage{graphicx}
\usepackage{epstopdf}

\begin{document}
\newcommand{\todo}[1]{{\em \small {#1}}\marginpar{$\Longleftarrow$}}   
\newcommand{\labell}[1]{\label{#1}\qquad_{#1}} 

\rightline{UPR-1221-T;  DCPT-10/55}   
\vskip 1cm 

\begin{center} {\Large \bf States of a chiral 2d CFT}
\end{center} 
\vskip 1cm   
  
\renewcommand{\thefootnote}{\fnsymbol{footnote}} 
\centerline{\bf Vijay Balasubramanian$^a$\footnote{vijay@physics.upenn.edu}, Jamie Parsons$^b$\footnote{j.d.parsons@durham.ac.uk} and
  Simon F. Ross$^b$\footnote{S.F.Ross@durham.ac.uk} }
\vskip .5cm 
\centerline{\it {}$^a$ Department of Physics \& Astronomy, University of Pennsylvania}
\centerline{\it 209 South 33rd St,  Philadelphia, PA USA}
\vskip .5cm
\centerline{\it {}$^b$ Centre for Particle Theory, Department of Mathematical Sciences}
\centerline{\it Durham University, South Road, Durham DH1 3LE, U.K.}

\setcounter{footnote}{0}   
\renewcommand{\thefootnote}{\arabic{footnote}}

\begin{abstract}
We study the dual description of the self-dual orbifold, a locally AdS$_3$ spacetime which is a circle fibration over AdS$_2$ and arises as the near-horizon limit of the extreme BTZ black hole. The geometry has two boundaries; we argue that this should correspond to a saddle-point for two copies of a chiral CFT living on these two boundaries in an entangled state. This picture arises naturally in the near-horizon limit, but there is a potential inconsistency with the bulk physics because of causal connections between the boundaries. We discuss a possible resolution of this puzzle. We also construct geometries  which asymptotically approach the self-dual orbifold on a single boundary. These geometries (which contain mild singularities) enable us to explore other states of the dual chiral CFT.  One of the geometries corresponds to the ground state of this CFT and can be obtained as a particular near-horizon limit of the BTZ $M=0$ black hole.   The self-dual orbifold is a finite temperature version of this geometry.
\end{abstract}

\section{Introduction}

There has recently been interest in the holographic relation between spacetime geometries with AdS$_2$ factors and dual field theories. One example is the proposed Kerr/CFT correspondence \cite{Guica:2008mu}, which attempts to extend holography to provide a description of the near-horizon region of uncharged extreme black holes. Another comes from studies of field theories at finite charge density in the AdS/CFT correspondence, which involve a Reissner-Nordstr\"om AdS black hole. This black hole has a near-horizon AdS$_2 \times \mathbb{R}^n$ geometry in the low-temperature limit, which controls the long-distance transport properties of the field theory \cite{Faulkner:2009wj}.    String theory in AdS$_2$ arises in the near-horizon limits of a wide variety of four and five dimensional black holes in both asymptotically flat (e.g., \cite{Strominger:1998yg}) and asymptotically AdS space (e.g., \cite{Balasubramanian:2007bs}).

The self-dual orbifold of AdS$_3$ \cite{Coussaert:1994tu} is a simple example of a geometry with an AdS$_2$ factor. This spacetime is a circle fibration over AdS$_2$, with an $SL(2,\mathbb{R}) \times U(1)$ isometry group, and can be viewed as arising either as a quotient of AdS$_3$, or as the near-horizon limit of a BTZ black hole \cite{sd1,sd2}. We can use these descriptions to understand the dual field theory description in detail. It has two asymptotic boundaries; from the quotient point of view, this is because the quotient has fixed points on the conformal boundary of AdS$_3$, and excising these fixed points divides the boundary into two disconnected regions.  The conformal geometry on these boundaries is a null cylinder; that is, a flat two-dimensional spacetime with a null direction compactified. Working in coordinates which only covered one boundary,
 it was argued  in \cite{sd1,sd2} (see Sec.~2) that the dual field theory is the Discrete Lightcone Quantization (DLCQ) of the original two-dimensional field theory dual to AdS$_3$, with the chiral sector which survives DLCQ in a thermal state (see also \cite{Strominger:1998yg}).  This raises a question:  {\it Are there any other states of the DLCQ theory that have a dual description as classical spacetimes that are asymptotic to the self-dual orbifold?}  
 
 In \cite{sd2}, the geometry was considered in the analogue of Poincar\'e coordinates, which only see one boundary of the spacetime, but the global self-dual orbifold has two disjoint boundaries \cite{sd1}. This raises a second long-standing question in the dual CFT description of asymptotically AdS$_2$ spacetimes:  {\it Are they dual to a single CFT, or to two copies of the CFT living on the two boundaries?}

In Sec.~\ref{global} we will argue that the self-dual orbifold should be thought of as dual to two copies of the CFT. We will first give a general argument based on the bulk diffeomorphism symmetries. We will then consider the orbifold global coordinates as coordinates on AdS$_3$ (without considering the quotient); here it is clear that there are independent CFT degrees of freedom on the two boundaries. Finally, we will observe that the self-dual orbifold in a ``black hole" coordinate system can be obtained by considering a near-horizon, near-extremal limit of the non-extremal BTZ  black hole. These coordinates cover regions of both boundaries. Since the non-extremal BTZ black hole is described by an entangled state in two copies of a CFT, in the near-horizon infrared limit we still have a pair of entangled CFTs (a similar argument was previously given in \cite{Azeyanagi:2007bj}).  However, the infrared limit restricts both theories to a chiral sector, and these sectors are entangled.  This is consistent with the picture of \cite{sd2}; tracing over one boundary will give us a thermal state in a chiral CFT.   We can also see the proposed entanglement by thinking of the orbifold global coordinates as coordinates on AdS$_3$: these cover the conformal boundary of AdS$_3$ in two patches, and rewriting the vacuum state of a CFT on the conformal boundary in these coordinates gives rise to entanglement of the right-moving degrees of freedom. This interpretation is also consistent with the general picture that the connectivity of regions of the boundary through the bulk is dual to entanglement between these regions in the CFT \cite{VanRaamsdonk:2010pw}. 

Extrapolating from this example suggests that the dual of any spacetime with an AdS$_2$ factor is a $1+1$ CFT with one chiral sector in its ground state, and the other  in an entangled state with non-zero entropy. The entanglement plays a crucial role in explaining the spacetime structure. This is quite different from the dual description of higher-dimensional AdS spaces, which correspond to ground states of a dual CFT, with no entropy.  

The global description of AdS$_2$ also leads to a puzzle -- the two boundaries  are {\it causally} connected, implying non-zero commutators for operators in the two copies of the CFT (Sec.~\ref{causal})\footnote{It is interesting to note that this problem is special to the two-dimensional case; in higher dimensions, \cite{Chrusciel:2008uz} shows that generically spacetimes cannot have disconnected boundaries that are causally connected through the bulk.}.  Entanglement cannot reproduce these commutators: if the two copies of the CFT are independent, the operators should commute.  Within the patches of AdS$_2$ that are recovered by the near-horizon limit of higher dimensional black holes, the puzzle is avoided because there is no causal connection between the regions of the two boundaries covered by the ``black hole'' coordinate systems.  The puzzle would be resolved if we were restricted to only consider correlation functions between spacelike separated operators on the two boundaries of AdS$_2$, perhaps in view of the AdS$_2$ instability of \cite{Maldacena:1998uz}.    Understanding whether such restrictions can be implemented is an important goal for future work.

The other main aim of our work is to construct new examples of asymptotically self-dual orbifold spacetimes (Sec.~\ref{asymp}) corresponding to different states of the dual field theory.  One motivation is to identify geometries dual to particular pure states that contribute to the entropy of the spacetime as in the black hole microstates program (see \cite{Mathur:2005zp} for a review).  Another motivation is that in the Kerr-CFT correspondence, it has been shown that there are no non-trivial asymptotically near-horizon extremal Kerr geometries \cite{Amsel:2009ev,Dias:2009ex}; it would be interesting to know if this is a general feature of spacetimes with AdS$_2$ factors.

We first show that there is a quotient of AdS$_3$ which is the natural dual to the ground state for a single copy of a CFT on a null cylinder, and which can be obtained as a near-horizon limit of the $M=0$ BTZ black hole.\footnote{This material, which appears in Sec.~4.2, has overlap with \cite{joanetal}.}   This is a quotient along a null direction in the bulk, so the geometry has closed null curves, and the duality may only be a formal correspondence in this case. This is similar to the Schr\"odinger spacetime \cite{Son:2008ye,Balasubramanian:2008dm}, where this issue was pointed out in \cite{Maldacena:2008wh}.  We then construct a rich class of  examples of asymptotically self-dual orbifold spacetimes by applying a solution-generating transformation \cite{Garfinkle:1990jq}. These geometries could be interpreted in the dual CFT as more restricted thermal ensembles where the local charge density is prescribed and not just the total charge. However, they are mildly singular in the bulk on the boundary of the region covered by the orbifold Poincar\'e coordinates. It is worth noting that this solution-generating approach relies on the existence of a globally null Killing vector, which appears in the self-dual orbifold but not in the higher-dimensional solutions with AdS$_2$ factors such as near-horizon extremal Kerr or the near-horizon limits of extreme Reissner-Nordstr\"om AdS black holes.

\section{The self-dual orbifold}
\label{review}

The self-dual orbifold spacetime was  introduced in \cite{Coussaert:1994tu} as a quotient of AdS$_3$, and its interpretation in the AdS/CFT correspondence was  discussed by \cite{sd1}.  The spacetime is a quotient along the Killing vector $\xi = U \partial_X + X \partial_U + V \partial_Y + Y \partial_V$, where $U, V, X, Y$ are the embedding coordinates which realise AdS$_3$ as a surface in $\mathbb{R}^{2,2}$. This Killing vector has $||\xi||^2 = 1$, so the quotient has no fixed points in the bulk. The quotient preserves an $SL(2,\mathbb{R}) \times U(1)$ subgroup of the $SL(2,\mathbb{R}) \times SL(2,\mathbb{R})$ symmetry of AdS$_3$, where the $U(1)$ factor is generated by $\xi$. There is a global coordinate system $(t,\phi,z)$ which covers the whole spacetime, related to the embedding coordinates by
\begin{align} \label{g2}
U+X &= \frac{1}{\sqrt{2}} e^\phi (e^z \cos t - e^{-z} \sin t), \\
U-X &= \frac{1}{\sqrt{2}} e^{-\phi} (e^{-z} \cos t - e^z \sin t) , \\
V+Y &= \frac{1}{\sqrt{2}} e^{\phi} (e^{-z} \cos t + e^z \sin t), \\
V-Y &= \frac{1}{\sqrt{2}} e^{-\phi} (e^z \cos t + e^{-z} \sin t). 
\end{align}
These coordinates are related to the usual global AdS$_3$ coordinates $(\rho,\tau,\theta)$ by
\begin{align} \label{ctgsd2}
\cosh^2 \rho &= \cosh^2 z \cosh^2 \phi + \sinh^2 z \sinh^2 \phi, \\
\tan (\tau+\theta) &= - \frac{\tanh 2z}{\sinh 2\phi}, \nonumber \\
\tan (\tau-\theta) &=  \frac{\tanh 2 \phi \cos 2t + \sinh 2z \sin 2t}{-\tanh
  2\phi \sin 2t + \sinh 2z \cos 2t}. \nonumber
\end{align}
in terms of which the metric of AdS$_3$ is 
\begin{equation} \label{sdglobal}
ds^2 = -dt^2 + d\phi^2 + 2 \sinh 2z dt d\phi + dz^2 = -\cosh^2 2z dt^2
+ (d\phi + \sinh 2z dt)^2 + dz^2 \, .
\end{equation}
The self-dual orbifold is obtained by taking the quotient $\phi \sim \phi + 2\pi r_+$ for some $r_+$.  The spacetime is then a $U(1)$ fibration over AdS$_2$, and the Killing symmetries are 
\begin{align}
\xi &= \frac{1}{2} \partial_\phi, \\
\chi_1 &= \frac{1}{2} \partial_t, \\
\chi_2 &= \frac{1}{2} \tanh 2z \cos 2t \partial_t + \frac{\cos 2t}{2
  \cosh 2z} \partial_\phi + \frac{1}{2} \sin 2t \partial_z, \\
\chi_3 &= -\frac{1}{2} \tanh 2z \sin 2t \partial_t - \frac{\sin 2t}{2
  \cosh 2z} \partial_\phi + \frac{1}{2} \cos 2t \partial_z.
\end{align}
The factors of 2 in the $\chi_i$ are required to make them a
representation of $SL(2,\mathbb{R})$; the one in $\xi$ is simply conventional. 

The spacetime has two boundaries, at $z \to \pm \infty$.  Taking (\ref{sdglobal}) as a coordinate system on all of AdS$_3$ (without a quotient) we would also have reached the boundary when $\phi \to \pm \infty$, which are fixed points of the quotient $\phi \sim \phi + 2\pi r_+$.   From \eqref{ctgsd2}, when $\phi \to \pm \infty$, 
\begin{equation}
\tan(\tau + \theta) = 0,
\end{equation}
so this corresponds to $\tau +\theta = 0$ or $\pi$. That is, the quotient has fixed points on the null lines $\tau+\theta =  0,\pi$ in the conformal boundary. These lines separate the conformal boundary into two strips. These two regions are the two boundaries of the self-dual orbifold, at $z \to \pm \infty$. When  $z \to \pm \infty$, \eqref{ctgsd2} simplifes to
\begin{equation}
\tan(\tau + \theta) = \mp \frac{1}{\sinh 2\phi}, \quad
\tan(\tau-\theta) = \tan 2t. 
\end{equation}
So on the boundary $t$ is mapped to the null coordinate running up along the strips at $z\to \pm \infty$, while $\phi$ is the null coordinate running across the strips. 

\begin{figure}
\begin{center}
\includegraphics[width=10cm]{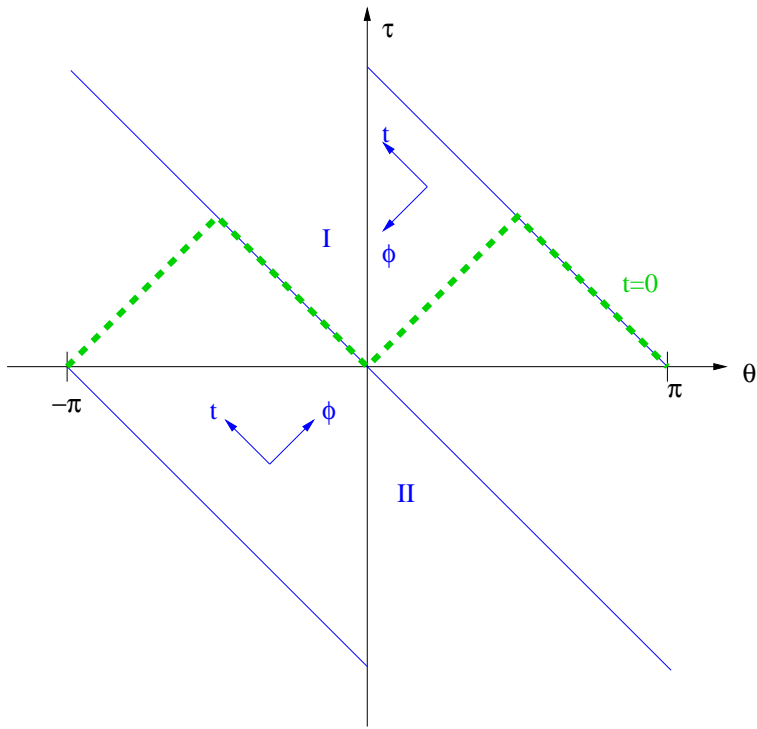}
  \caption{The relation between $\tau,\theta$ and $t,\phi$ coordinates on the boundary. (The lines $\theta = \pm \pi$ are identified.) Region I is 
$z = \infty$, region II is $z = -\infty$. The direction of increasing $t, \phi$ in each region is 
indicated. The heavy dashed line is the surface $t=0$ given in \eqref{g2b}.}
\label{fig:orb1}
\end{center}
\end{figure}

Consider a surface of constant $t$, say $t=0$. In the
strips at $z = \pm \infty$, this will map to $\tau-\theta = 0,
\pi$. Let's choose $\tau-\theta = 0$ at $z = \infty$. At $t=0$, 
\begin{equation}
\tan \theta = -\frac{\tanh \phi - \tanh z}{\tanh \phi + \tanh z}. 
\end{equation}
At $z = \infty$, as $\phi$ ranges from $-\infty$ to $\infty$, $\theta$
ranges over $(\pi/2,0)$. At $\phi = \infty$, as $z$ ranges from
$\infty$ to $-\infty$, $\theta$ ranges over $(0,-\pi/2)$. At $z =
-\infty$, as $\phi$ ranges from $\infty$ to $-\infty$, $\theta$ ranges
over $(-\pi/2, \pi)$. Finally, at $\phi = -\infty$, as $z$ ranges from
$-\infty$ to $\infty$, $\theta$ ranges from $(\pi,\pi/2)$. As a
result, the surface $t=0$ maps to a sawtoothed curve made from null
segments: 
\begin{equation}
t=0 \leftrightarrow \left\{ \begin{array}{cc} \tau -\theta = 0, & \theta \in (\pi/2, 0), \\ \tau + \theta = 0,& 
\theta \in (0, -\pi/2), \\ \tau - \theta = \pi,& \theta \in (-\pi/2, -\pi) \\ \tau + \theta = \pi,& \theta \in (\pi, \pi/
2). \end{array} \right. 
\end{equation}
This is shown in figure \ref{fig:orb1}.

In addition to arising as a quotient of AdS$_3$, the self-dual orbifold can be obtained as a near-horizon limit of the extremal BTZ black hole, a point of view which was stressed in \cite{sd2}. If we start with the BTZ black hole in a stationary coordinate system, we obtain the self-dual orbifold in a coordinate system which only covers a portion of the geometry. A convenient coordinate system is the $(u,v,r)$ coordinates introduced in \cite{sd2}, which are related to the embedding coordinates by
\begin{align} \label{g2b}
U+X &=  e^{r+u}, \\
U-X &=  \frac{1}{2}
(e^{-r-u} -2 v e^{r-u}), \\
V+Y &=  \frac{1}{2}
(e^{-r+u} +2 v e^{r+u}), \\
V-Y & = e^{r-u}. 
\end{align}
The transformation between the $(t,\phi,z)$ and $(u,v,r)$ coordinates is then
\begin{align}
2 e^{2r} &= \cosh 2z \cos 2t + \sinh 2z, \label{ad2t1} \\
v &= \frac{ \cosh 2z \sin 2t}{ \cosh 2z \cos 2t + \sinh
  2z}, \label{ads2t2} \\
e^{2u} &= e^{2 \phi} \frac{(e^z \cos t - e^{-z} \sin t)}{(e^z \cos t +
  e^{-z} \sin t)}.
\end{align}
These coordinates are also simply related to the Poincar\'e coordinates $(x^+, x^-, Z)$ on AdS$_3$,
\begin{equation} \label{Poinc}
Z = e^{-r+u}, \quad x^+ = \frac{1}{2} e^{-2u}, \quad  x^- = v - \frac{1}{2} e^{-2r}. 
\end{equation}

The metric of AdS$_3$ in this coordinate system is 
\begin{equation} \label{sdp}
ds^2 = du^2 + 2e^{2r} du dv + dr^2 = - e^{4r} dv^2 + dr^2 +
(du+ e^{2r} dv)^2,
\end{equation}
and the Killing vectors are
\begin{align} \label{PKV}
\xi &= \frac{1}{2} \partial_u, \\
\chi_1 &= -\left[ 1 + \frac{1}{4} (v^2+ e^{-4r}) \right] \partial_v -
\frac{1}{4} e^{-2r} \partial_u + \frac{1}{4} v \partial_r , \\
\chi_2 &=  -\left[ 1 - \frac{1}{4} (v^2+ e^{-4r}) \right] \partial_v +
\frac{1}{4} e^{-2r} \partial_u - \frac{1}{4} v \partial_r \\
\chi_3 &= - v \partial_v + \frac{1}{2} \partial_r. 
\end{align}
The $(t,\phi, z)$ coordinate system covers the whole of AdS$_3$, as do the $(\tau, \theta, \rho)$ global coordinates, but from \eqref{Poinc}, we can see that $(u,v,r)$ coordinates cover half a Poincar\'e patch of AdS$_3$, as they cover the region $x^+ >0$.   

From the AdS$_2$ point of view, the transformation (\ref{ad2t1},\ref{ads2t2}) is precisely the transformation from global coordinates to Poincar\'e coordinates on the AdS$_2$ factor, so these coordinates will cover the region of the self-dual orbifold corresponding to the Poincar\'e patch in AdS$_2$. It is interesting that half of the Poincar\'e patch in AdS$_3$ corresponds to the Poincar\'e patch in AdS$_2$. This patch includes a portion of the boundary at $z = \infty$ in the global coordinates $(t,\phi,z)$. The other half of the AdS$_3$ Poincar\'e patch corresponds to an AdS$_2$ Poincar\'e patch covering a portion of the other boundary at $z=-\infty$. 

The coordinate transformation simplifies further on the boundary. At $z \to \infty$, 
\begin{equation}
u = \phi, \quad v =  \tan t.
\end{equation}
so the $(u,v)$ coordinates cover the portion of the $z=\infty$ strip
with $t \in (-\pi/2,\pi/2)$.  The relation between Poincar\'e coordinates $x^+, x^-$ and the $u, v$ coordinates on the boundary is
\begin{equation} \label{Poincbd}
x^+ = \frac{1}{2} e^{-2u}, \quad x^- =  v.
\end{equation}

The self-dual orbifold spacetime can be obtained as a near-horizon limit of an extreme BTZ black hole. In \cite{sd2}, it was observed that this near-horizon limit is very easy to describe in the $(u,v,r)$ coordinates \eqref{sdp}. The extreme BTZ black hole is given by this metric with the identifications 
\begin{equation}
(\tilde u, \tilde v) \sim (\tilde u + 2 \pi r_+, \tilde v + 2\pi r_+).
\end{equation}
We call the extreme BTZ coordinates $(\tilde u, \tilde v, \tilde r)$ to distinguish them from the self-dual orbifold coordinates we will shortly recover. To take the near-horizon limit, we want to write $\tilde r = r_0 +  r$, and take $r_0 \to -\infty$. If we also write $\tilde u=  u$, $\tilde v= e^{-2r_0} v$, then the metric in terms of $(u, v, r)$ takes the same form \eqref{sdp} at finite $r_0$, but now with the identifications  
\begin{equation} \label{did}
(u, v) \sim (u + 2 \pi r_+, v + 2\pi r_+ e^{2r_0}).
\end{equation}
As we take the near-horizon limit $r_0 \to -\infty$ for fixed $(u,v,r)$, this reduces to $u \sim u+2\pi r_+$, giving us the self-dual orbifold. 

The identifications \eqref{did} describe a circle of proper size $2 \pi r_+ e^{r_0}$ viewed in a boosted frame. So taking this limit corresponds to a DLCQ limit in the CFT on the $(u,v)$ cylinder,  where we take the size of the circle to zero and the boost to infinity to recover a null identification, giving the CFT on a null cylinder. In general, this DLCQ limit will restrict us to the ground state for the right-moving excitations, as the energy of right-moving excitations is infinitely blueshifted by the boost: $E_v = E_{\tilde v} e^{-2 r_0}$. We might think that we could take $E_{\tilde v} \to 0$ at the same time to recover a finite energy in the boosted frame, but since the theory lives in finite volume, there is a finite density of states, and the energy spectrum is quantised. The spacing is at least $\Delta E_{\tilde v} \sim e^{-S} \sim e^{-c}$, where $c$ is the central charge. Although this discreteness in the spectrum cannot be seen from the classical spacetime point of view, it will ensure that we are ultimately left with only the ground states for the right-movers, with  $L_0 = \frac{c}{24}$. The dual CFT is thus a chiral theory, with only left-moving excitations. 

The left-moving excitations are unaffected by this DLCQ procedure, so we would expect them to be in the same state as before we took the near-horizon limit. The dual of the extreme BTZ black hole is a thermal state for the left-movers, at temperature $T_L = r_+/2\pi$. Since we have included a factor of $r_+$ in our definition of the identifications, the temperature with respect to our coordinate $u$ is in fact $T = 1/2\pi$. 

Thus, the proposal of \cite{sd2} was that the dual description was a chiral CFT, with the left-movers in a thermal state at temperature $T = 1/2\pi$. In section \ref{global}, we will refine this proposal by considering the description of the self-dual orbifold including both boundaries, and propose that the dual is an entangled state which reduces to this thermal state on tracing over the CFT degrees of freedom on one of the boundaries.

\subsection{Black hole coordinates}
\label{bhc}

The self-dual orbifold spacetime has two boundaries, but when we obtain it as a near-horizon limit of the extreme BTZ black hole, we obtain it in a coordinate system which only covers one of the boundaries. To see that both boundaries play a role, it is useful to consider a different kind of near-horizon limit. We therefore consider the near-horizon, near-extremal limit of the non-extremal BTZ black hole. 

We start with the BTZ black hole,
\begin{equation} \label{btz}
ds^2 = - \frac{(r^2-r_+^2)(r^2 - r_-^2)}{r^2} dt^2 + \frac{r^2
  dr^2}{(r^2 -r_+^2) (r^2-r_-^2)} + r^2 \left( d\phi - \frac{r_-
    r_+}{r^2} dt \right)^2, 
\end{equation}
which has two asymptotic regions in the full eternal black hole spacetime. The coordinate range $r \geq 0$ only covers one asymptotic boundary, but the metric can be extended in patches to cover both boundaries.

We 
first define a comoving coordinate system at the event horizon $r=r_+$
by setting $\phi' = \phi - \frac{r_-}{r_+} t$. Then define new
coordinates $(\bar t, \bar \phi, \bar r^2)$  by
\begin{equation}
r^2 = r_+^2 (1 + \epsilon \bar r^2), \quad t = \frac{\bar t}{r_+
  \epsilon}, \quad \phi' = \frac{\bar \phi}{r_+}, 
\end{equation}
where $\epsilon = \frac{r_+^2 - r_-^2}{r_+^2}$. The metric in these
coordinates is
\begin{equation}
ds^2 = - \frac{\bar r^2 (\bar r^2+1)}{(1 + \epsilon \bar r^2)} d \bar
t^2 + \frac{ (1 + \epsilon \bar r^2)}{(\bar r^2 + 1)} d \bar r^2 + (1
+ \epsilon \bar r^2) \left( d \bar \phi + \frac{\sqrt{1-\epsilon^2}}{(1
+ \epsilon \bar r^2)} \bar r^2 d \bar t\right)^2.  
\end{equation}
We can then take a near-horizon, near-extremal limit by taking 
$\epsilon \to 0$ for finite values of $(\bar t, \bar
\phi, \bar r^2)$. The resulting metric is 
\begin{equation} \label{bhmet}
  ds^2 = - \bar r^2 (\bar r^2+1) d \bar
  t^2 + \frac{ d \bar r^2}{(\bar r^2 + 1)} + \left( d \bar \phi + \bar
    r^2 d \bar t\right)^2 = -\bar r^2 d \bar t^2 + d \bar \phi^2 + 2
  \bar r^2 d \bar t d \bar \phi + \frac{ d \bar r^2}{(\bar r^2 + 1)}.   
\end{equation}
This is the self-dual orbifold, in a ``black-hole like'' coordinate system. The AdS$_2$ part of the metric is written in the black hole coordinates of eq (2.12) of \cite{Maldacena:1998uz}. It is also worth noting that the inner horizon at $\bar r^2 = -1$ remains at a finite distance from the outer horizon at $\bar r^2 = 0$ as we take this limit. 

To relate these coordinates to embedding coordinates, take the embedding of BTZ,
\begin{align}
U &= \sqrt{ \frac{r^2-r_+^2}{r_+^2 - r_-^2} } \sinh (r_+ t - r_- \phi)
= \sqrt{ \frac{r^2-r_+^2}{r_+^2 - r_-^2} } \sinh \left( \frac{r_+^2 -
  r_-^2}{r_+}  t - r_- \phi' \right),\\
V &= \sqrt{ \frac{r^2-r_-^2}{r_+^2 - r_-^2} } \cosh (r_+ \phi - r_- t)
= \sqrt{ \frac{r^2-r_+^2}{r_+^2 - r_-^2} } \cosh (r_+ \phi'),
\nonumber \\
X &= -\sqrt{ \frac{r^2-r_+^2}{r_+^2 - r_-^2} } \cosh (r_+ t - r_- \phi)
= -\sqrt{ \frac{r^2-r_+^2}{r_+^2 - r_-^2} } \cosh \left( \frac{r_+^2 -
  r_-^2}{r_+}  t - r_- \phi' \right), \nonumber \\
Y &= \sqrt{ \frac{r^2-r_-^2}{r_+^2 - r_-^2} } \sinh (r_+ \phi - r_- t)
= \sqrt{ \frac{r^2-r_+^2}{r_+^2 - r_-^2} } \sinh (r_+ \phi'), \nonumber
\end{align}
and apply the same $\epsilon \to 0$ limit. This gives
\begin{align}
U+X &= - \bar r e^{-\bar t + \bar \phi}, \\
U-X &= \bar r e^{\bar t - \bar \phi}, \nonumber \\
V+Y &= \sqrt{ \bar r^2 +1} e^{\bar \phi}, \nonumber \\
V-Y &= \sqrt{ \bar r^2 +1} e^{-\bar \phi}.\nonumber 
\end{align}
These coordinates are related to the global $(t,\phi,z)$ coordinates by 
\begin{align} 
\bar r^2 &= \sinh^2 z \cos^2 t - \cosh^2 z \sin^2 t, \\
e^{2 \bar t} &= \frac{\tanh 2z + \sin 2t}{\tanh 2z - \sin 2t} \nonumber
\\ 
e^{2 \bar \phi} &= e^{2 \phi} \frac{(\cosh z \cos t + \sinh z \sin
  t)}{(\cosh z \cos t - \sinh z \sin t)}. \nonumber
\end{align}
Thus, the $(\bar t, \bar \phi, \bar r)$ coordinates for $\bar r^2
\geq 0$ cover a region $z \geq 0$, $\tan^2 t \leq \tanh^2 z$ in
$(t,\phi,z)$ coordinates.    Just as the non-extreme BTZ black hole has two asymptotic regions, which can be displayed by taking two patches with metrics (\ref{btz}), we can think of the black hole coordinates for the self-dual orbifold as covering  two regions outside the ``horizon'' at $\bar r^2=0$, thus including patches of the two boundaries at $z \to \pm \infty$. 

Note that in these black hole coordinates, $\bar t, \bar \phi$ are not
null coordinates on the boundary of the self-dual orbifold. This can
be corrected by defining $\tilde \phi = \bar \phi - \frac{\bar t}{2}$;
it's also useful to set $\tilde t = \frac{\bar t}{2}$.  Then the
metric in these coordinates is
\begin{equation} \label{bhmet2}
  ds^2 = - 4 \bar  r^2 (\bar r^2+1) d \tilde
  t^2 + \frac{ d \bar r^2}{(\bar r^2 + 1)} + \left( d \tilde \phi + 
     ( 2\bar r^2 + 1 ) d \tilde t\right)^2, 
\end{equation}
and the relation to $(t,\phi,z)$ coordinates is 
\begin{align} 
\bar r^2 &= \sinh^2 z \cos^2 t - \cosh^2 z \sin^2 t, \\
e^{4 \tilde t} &= \frac{\tanh 2z + \sin 2t}{\tanh 2z - \sin 2t} \nonumber
\\ 
e^{4 \tilde \phi} &= e^{4 \phi} \frac{\sinh 2z - \tan 2t}{\sinh 2z +
  \tan 2t}. \nonumber
\end{align}

The Killing vectors in these coordinates are
\begin{align}
\xi =& \frac{1}{2} \partial_{\tilde \phi}, \\
\chi_1 =& -\frac{1}{4} \left( \frac{\sqrt{\bar r^2 +1}}{\bar r}
  +\frac{\bar r} {\sqrt{\bar r^2 +1}} \right) \cosh 2 \tilde
t \partial_{\tilde t} + \frac{1}{4} \left( \frac{\sqrt{\bar r^2 +1}}{\bar r}
  -\frac{\bar r} {\sqrt{\bar r^2 +1}} \right) \cosh 2 \tilde
t \partial_{\tilde \phi} \nonumber  \\ &+ \sqrt{\bar r^2 +1} \sinh 2 \tilde
t \partial{\bar r},   \\
\chi_2 =&  \frac{1}{4} \left( \frac{\sqrt{\bar r^2 +1}}{\bar r}
  +\frac{\bar r} {\sqrt{\bar r^2 +1}} \right) \sinh 2 \tilde
t \partial_{\tilde t} - \frac{1}{4} \left( \frac{\sqrt{\bar r^2 +1}}{\bar r}
  -\frac{\bar r} {\sqrt{\bar r^2 +1}} \right) \sinh 2 \tilde
t \partial_{\tilde \phi}\nonumber \\ &- \sqrt{\bar r^2 +1} \cosh 2 \tilde
t \partial{\bar r},  \\
\chi_3 =& - \frac{1}{2} \partial_{\tilde t}. 
\end{align}

On the boundary at $z \to \infty$,
\begin{equation} \label{bhct}
e^{2 \tilde \phi} = e^{2 \phi},
\quad e^{2 \tilde t} = \frac{\cos t + \sin t}{\cos t - \sin t};
\end{equation}
note that the rescaling of $\tilde t$ was chosen so that for $t$ near
$0$, $\tilde t \approx t$. This coordinate system covers the portion
of the $z=\infty$ strip with $t \in (-\pi/4,\pi/4)$; half as much as
the $u,v$ coordinates. In terms of the $u,v$ coordinates,
\begin{equation} \label{bhctv}
e^{2 \tilde \phi} = e^{2 \phi},
\quad e^{2 \tilde t} = \frac{1 + v}{1 - v},
\end{equation}
so it maps to the region $v \in (-1,1)$.

\begin{figure}
\begin{center}
\includegraphics[height=3cm]{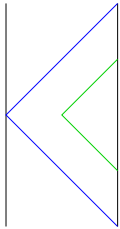}
  \caption{The regions in AdS$_2$ covered by the different coordinate
    systems. The $(t,z)$ coordinates cover the whole of AdS$_2$, the
    $(v,r)$ coordinates cover the Poincar\'e wedge indicated in blue,
    and the $(\bar t, \bar r)$ coordinates cover the smaller wedge
    indicated in green.}
\label{fig:orb2}
\end{center}
\end{figure}

\section{Holographic dual of the global spacetime}
\label{global}

The self-dual orbifold has two boundaries. We would like to understand whether this should be interpreted as the dual of a pair of field theories on these two boundaries, or whether there should be a single CFT dual to the spacetime, as has been proposed for AdS$_2$ in \cite{Strominger:1998yg}. We will argue from several points of view that the correct interpretation seems to be as an entangled state in two disjoint copies of the CFT, living on the two boundaries of the spacetime.  We will see that a challenge to this interpretation arises because the two boundaries of the self-dual orbifold are causally connected, suggesting that there is an interaction between them.

\subsection{Diffeomorphisms and Hamiltonians}

 There is a simple general argument using the bulk diffeomorphism symmetry which implies that the dual description should indeed involve two copies of the field theory. Consider a spacetime with two conformal boundaries. If we consider a bulk surface of constant $t$, the bulk diffeomorphism freedom allows us to shift the surface arbitrarily. Changes in the constant $t$ surface which do not affect its intersection with the boundary are pure gauge, and are not seen from the boundary field theory point of view. But diffeomorphisms which shift the intersection with the boundary correspond to the action of a boundary Hamiltonian from the field theory point of view \cite{Marolf:2008mf, Marolf:2008mg}. Since the bulk diffeomorphism symmetry includes transformations which independently deform the intersection of the bulk constant $t$ surface with the two boundaries, there are two such Hamiltonians, acting separately on the two boundaries. These generate a $\mathbb{R} \times \mathbb{R}$ symmetry of the theory, corresponding to arbitrary time translations in the two CFTs.\footnote{Actually, since the bulk surface meets the boundary in a one-dimensional surface in the self-dual orbifold, there is an infinite-dimensional group of translations of this surface in each boundary, but let us restrict for simplicity to the overall translational subgroup.}  We take the presence of these two independent time-translation generators to imply that there are two separate physical systems on the two boundaries. If the bulk geometry corresponded in the dual theory to a state in a single field theory, we would expect to have just one Hamiltonian, not two. This argument is quite general and would apply whenever the spacetime has two boundaries. 

In the self-dual orbifold, the bulk spacetime has a time-translation symmetry that corresponds to the diagonal $\mathbb{R}$ subgroup of the $\mathbb{R} \times \mathbb{R}$ symmetry of the theory. This implies that the dual CFT description should be in terms of a state in the two CFTs which preserves this diagonal $\mathbb{R}$ subgroup. This state should involve some non-trivial entanglement between the two theories, as the connectedness of the spacetime in the bulk implies that there will be non-zero correlation functions $\langle \mathcal O_1 \mathcal O_2 \rangle$ between operators on the two boundaries. This relation between entanglement and connections in the bulk spacetime was investigated in general in \cite{VanRaamsdonk:2010pw}. We will work out the form of the entanglement in the case of the self-dual orbifold a little further on.

\subsection{Quotient perspective}

To understand the holographic description in more detail, we consider the spacetime from the quotient perspective. As a first step, we can consider the metric \eqref{sdglobal} with $\phi$ non-compact as simply a new choice of coordinates on AdS$_3$, where we understand holography well.

From this perspective, it is clear that there are independent CFT degrees of freedom in the two boundaries, as these correspond to different regions in the single boundary of AdS$_3$. The two Hamiltonians found above correspond to translations of the two segments of the $t=0$ surface in figure \ref{fig:orb1}. 

If we take a linear quantum theory on the boundary as a toy model for the CFT on the cylinder, we can use the coordinate transformations given above to work out the description of its vacuum state in terms of the self-dual orbifold coordinates.\footnote{
This perspective was used to identify the state dual to the non-extremal BTZ black hole in\cite{Maldacena:1998bw}; this was extended to the extremal black hole in \cite{Marolf:2005ry}.}
It is most straightforward to do this using the transformation \eqref{Poincbd} between the $(x^+,x^-)$ Poincar\'e coordinates on AdS$_3$ and the $(u,v)$ coordinates. There is a non-trivial coordinate transformation between $x^+$ and $u$, so the ground state for left-movers with respect to $x^+$ will map to an entangled state where the modes on one boundary are entangled
with the corresponding modes on the other. Tracing over one boundary
will then leave us with a thermal state for the left-movers on the
other boundary. 

If we consider a massless scalar field on the
boundary, the positive frequency left and right-moving mode solutions
in $(u,v)$ coordinates are, up to normalisation,
\begin{equation}
p^1_{\omega,l} = e^{-i \omega u}, \quad p^1_{\omega,r} = e^{- i \omega
  v},
\end{equation}
\begin{equation}
p^2_{\omega,l} = e^{i \omega u}, \quad p^2_{\omega,r} = e^{-i \omega
  v},
\end{equation}
where $1,2$ denote the two $(u,v)$ coordinate patches, each of which
covers half of the Poincar\'e patch. In global coordinates, these two
patches lie on the boundaries at $z = \pm \infty$ respectively.  The
subscripts $l,r$ denote left and right-moving modes. We can construct
modes which are purely positive frequency with respect to Poincar\'e
coordinates by rewriting these solutions in terms of $x^+, x^-$ and
analytically continuing in the lower-half complex $x^\pm$
plane, following \cite{Unruh:1976db}. This is a standard exercise; the only difference here is that the
relation is only non-trivial for the left-movers. Using the
coordinate transformation $x^+ =  \frac{1}{2} e^{-2u}$, we see that solutions which
are pure positive frequency with respect to $x^+$ are 
\begin{equation}
W^1_{\omega,l} = p^1_{\omega,l} + e^{-\frac{\pi \omega}{2}} \bar
p^2_{\omega,l}, 
\end{equation}
and similarly analytically continuing the solution $p^2_{\omega,l}$ will give 
\begin{equation}
W^2_{\omega,l} = p^2_{\omega,l} + e^{-\frac{\pi \omega}{2}} \bar
p^1_{\omega,l}. 
\end{equation}
These are purely positive frequency modes with respect to $x^+$, so
the corresponding annihilation operators $a^1_{\omega,l},
a^2_{\omega,l}$ will annihilate the ground state. Using the above
expressions, we can write these annihilation operators in terms of the
annihilation operators $b^1_{\omega,l}, b^2_{\omega,l}$ for the modes
$p^1_{\omega,l}$, $p^2_{\omega,l}$ as
\begin{align}
a^1_{\omega,l} &= b^1_{\omega,l} - e^{-\frac{\pi \omega}{2}}
b^{2\dagger}_{\omega,l}, \\
a^2_{\omega,l} &= b^2_{\omega,l} - e^{-\frac{\pi \omega}{2}}
b^{1\dagger}_{\omega,l}.
\end{align}
This indicates that the vacuum $|0\rangle$ in Poincar\'e
coordinates can be formally written as an entangled state in the
Hilbert space built on the vacua $|0\rangle_1$, $|0\rangle_2$
annihilated by the $b^1_{\omega,l}, b^2_{\omega,l}$:
\begin{equation} \label{estate}
| 0 \rangle = e^{-i \int_0^\infty d\omega e^{-\pi \omega}
  b^{1\dagger}_{\omega,l} b^{2\dagger}_{\omega,l}} |0\rangle_1 \otimes
|0\rangle_2. 
\end{equation}
This demonstrates that from the point of view of self-dual orbifold
coordinates, the vacuum on the boundary of AdS$_3$ is an
entangled state where the left-movers on the two boundaries of the
self-dual orbifold are entangled with each other. If we trace over one boundary, we would have a thermal state for the left-movers on the other boundary, at a temperature $1/2\pi$. This is consistent with the analysis of \cite{sd2} because we have hidden the
scale in $u$; the quotient identification is $u \sim u + 2\pi r_+$.  

We should next consider the effect of the quotient. The quotient will have three effects; first, the points at $\phi = \pm \infty$, corresponding to $\tau +\theta = 0, \pi$, are fixed points of the quotient. We must therefore excise them from the boundary manifold, turning it into two genuinely disconnected surfaces. Secondly, if we write the metric on one of these surfaces in terms of the $(u,v)$ coordinates, the effect of the quotient is then to make $u$ periodic. This will restrict the momentum in the $u$ direction to discrete values, leaving us with the same entangled state, but with the integral in \eqref{estate} replaced by a sum. This reduces the $SL(2,\mathbb{R}) \times SL(2,\mathbb{R})$ symmetry of the state to $SL(2,\mathbb{R}) \times U(1)$. 

Thirdly, the quotient will project out the right-movers. This cannot be seen directly by imposing the quotient on the boundary, as the left movers have no $u$ dependence. However, we can think of this null identification as a limit of a spacelike identification, by thinking of the spacetime as cutoff at some finite $r$, and taking the limit $r \to \infty$ will give us an infinite boost. Thus, in this quotient perspective we can also see the DLCQ of the field theory that was seen in the near-horizon limit in \cite{sd2}.  The infinite boost as we take the limit $r \to \infty$ sets the right-movers to the ground state. For a simple scalar field model, this is easy to see:  the metric on a surface at $r=r_0$ is given by
\begin{equation}
ds^2 = du^2 + 2e^{2r_0} du dv.
\end{equation}
Thus the null coordinates at finite $r_0$ are $\tilde u = u$, $\tilde v = v - 2e^{-2 r_0} u$.  Right moving modes are $e^{i\omega_R \tilde v}$, whilst left moving modes are $e^{i\omega_L \tilde u}$. Looking at the right movers, the modes which survive the quotient are those with $\omega_R = n e^{2 r_0}$ for $n \in \mathbb{Z}$. Therefore as the cutoff is removed, $r_0 \to \infty$ and the only right mover remaining is the ground state, $n=0$.  For the actual CFT on the boundary, the picture is a little more subtle; the theory lives in finite volume, so the energy spectrum is quantised, but as was observed in our review of the DLCQ argument of \cite{sd2}, the level spacing is $\Delta L_0 \sim e^{-c}$. This discreteness cannot be seen from the spacetime point of view. Nonetheless, as the boost is taken to infinity, we need to take $L_0 - \frac{c}{24} \to 0$, and the energy will ultimately become smaller than this gap, forcing us to set $L_0 = \frac{c}{24}$.

This discussion considered a simple toy model of a single scalar field, but the key point is just the non-trivial transformation between $x^+$ and $u$, so the $SL(2,\mathbb{R}) \times SL(2,\mathbb{R})$ invariant vacuum state for the full boundary CFT will similarly be an entangled state for the left-movers on the two regions $z = \pm \infty$.  The fact that the CFT is in a non-trivial excited state can be seen directly from the fact that the boundary stress tensor takes a non-vanishing value. This can be calculated from the bulk metric \eqref{sdp} using the usual holographic dictionary \cite{Balasubramanian:1999re,Skenderis:2002wp}. The boundary stress tensor is 
\begin{equation}
T_{\alpha\beta} = \frac{c}{6}[ \pi_{\alpha \beta} + h_{\alpha\beta}]  = \frac{c}{6} [K_{\alpha\beta} - (K-1) h_{\alpha\beta}],
\end{equation}
where $K_{\alpha\beta}$ is the extrinsic curvature, $h_{\alpha\beta}$ is the induced metric on the boundary, and $c = 3/2G_3$ is the central charge of the boundary CFT. The metric \eqref{sdp} gives $K_{uv} = e^{2r}$, so $K=2$, and the only non-vanishing component of $T_{\alpha\beta}$ is
\begin{equation}
T_{uu} = -\frac{c}{6}.
\end{equation}
Note that this is a momentum density in coordinates where the $u$
direction is periodic with period $2\pi r_+$; the total left-moving
momentum on the boundary is hence $c \pi r_+/3$. 

This stress tensor could also be obtained by considering the Schwarzian derivative associated with the coordinate transformation \eqref{Poincbd} between the Poincar\'e coordinates $x^+$, $x^-$ and the  $u, v$ coordinates. In terms of  holomorphic coordinates $w,\bar{w}$, the Schwarzian derivative gives in general
\begin{equation} \label{sder}
T(w') =  T(w) (\partial_{w'} w)^2 + \frac{c}{12} \left[ \frac{\partial^3_{w'}
    w}{\partial_{w'} w} - \frac{3}{2} \left(
    \frac{\partial^2_{w'}w}{\partial_{w'} w} \right)^2 \right], 
\end{equation}
and similarly for $\bar{T}(\bar{w})$. Since the stress tensor in
Poincar\'e coordinates vanishes in the global vacuum state on the
boundary cylinder, and the coordinate transformation for the
left-moving coordinates is trivial, this implies that $T_{vv} = 0$,
and
\begin{equation}
T_{uu} = \frac{c}{12} \left[ \frac{\partial_u^3 x^+}{\partial_u x^+} -
  \frac{3}{2}  \left(
    \frac{\partial^2_{u}x^+}{\partial_{u} x^+} \right)^2 \right] =
  \frac{c}{12} [ 4 - 6] = - \frac{c}{6},
\end{equation}
reproducing the direct result. Thus, from the quotient point of
view, the boundary stress tensor is accounted for by the non-trivial
conformal transformation from the Poincar\'e coordinates to the
self-dual orbifold coordinates $(u,v)$, and is associated with the fact that the field theory is in a non-trivial entangled state on the two boundaries. 

Thus, the self-dual orbifold is identified with a non-trivial entangled state of two copies of the CFT, living on the two boundaries of the spacetime. The fact that the dual description of a spacetime with an AdS$_2$ factor involves an excited state is quite different from the description of higher-dimensional AdS spacetimes, which are usually dual to the vacuum state in the dual CFT. This description of AdS$_2$ is more analogous to the description of black holes in higher-dimensional AdS spacetimes.

However, we would argue that the description of AdS$_2$ will always be qualitatively similar to this. There should be some entanglement to account for the connectivity between the two boundaries in the AdS$_2$ spacetime  \cite{VanRaamsdonk:2010pw}, and geometries with an AdS$_2$ factor are usually obtained as the near-horizon limit of black holes with a non-zero entropy, which is reproduced by the entropy of the mixed state obtained by tracing over one of the boundaries.

\subsection{Near-horizon near-extremal limit}

We can also argue for this description of the self-dual orbifold by taking the non-extremal BTZ black hole and considering the near-horizon, near-extremal limit introduced in section \ref{bhc}. The dual description of the non-extremal BTZ black hole is as a saddle-point corresponding to an entangled state of two copies of the CFT on the two boundaries of the maximal analytic extension of the black hole,\footnote{This description was obtained by a similar quotient argument in \cite{Maldacena:1998bw} and by considering the analytic continuation from the Euclidean black hole in \cite{Maldacena:2001kr}.} with the left- and right-movers at temperatures 
\begin{equation}
T_L = \frac{(r_+ +r_-)}{4 \pi}, \quad T_R = \frac{(r_+ -r_-)}{4 \pi}.
\end{equation}
Taking the near-horizon, near-extremal limit of the geometry is a DLCQ limit from the point of view of the field theory. This DLCQ limit does not affect the structure of the field theory, so it will give us two copies of the CFT in an entangled state on the regions of the boundary of the self-dual orbifold covered by the black hole coordinates \eqref{bhmet2}.  This confirms the entanglement description of the self-dual orbifold geometry. 

However, there is a small subtlety in the nature of the entangled state in these coordinates. Naively one would say that as we take the extremal limit, $T_R \to 0$, but $T_L$ remains finite, reproducing the entangled state we saw above in $(u,v)$ coordinates. However, $T_R$ is the temperature with respect to the null coordinate $x^-$ in the boundary of the black hole spacetime; this is related to the right-moving coordinate $\tilde t$ in the near-horizon region by an infinite boost. Taking this boost into account, the temperature with respect to $\tilde t$ is
\begin{equation}
T_{\tilde t} = \frac{T_R}{r_+ \epsilon} = \frac{1}{2 \pi}, 
\end{equation}
while the temperature with respect to $\tilde \phi$ is $T_{\tilde \phi} = 1/2\pi$, as it was in the analysis above from the quotient point of view. Thus, when we consider the spacetime in black hole coordinates, it appears to be dual to an entangled state where both the left and right-movers are at finite temperature. This can also be seen by considering the boundary stress tensor in these coordinates, which is given by 
\begin{equation} \label{bhstress}
T_{\tilde \phi \tilde \phi} = -\frac{c}{6}, \quad T_{\tilde t \tilde t} = -\frac{c}{6}.
\end{equation}
This appears to be inconsistent with the statement that $L_0 = \frac{c}{24}$, which should follow from the DLCQ here as it did in \cite{sd2}. However, the two statements are in fact perfectly consistent.  In general, for the CFT on a spacelike circle, the translation generators are
\begin{equation}
L_0 - \frac{c}{24} = \oint T_{z \mu} n^\mu, \quad \bar L_0 -
\frac{c}{24} = \oint T_{\bar z \mu} n^\mu, 
\end{equation}
where the integral is around the spacelike circle, and $n^\mu$ is the
unit normal to this circle in the boundary metric. Since $T_{z \bar z} = 0$
for a spacelike circle, this reduces to 
\begin{equation}
L_0 - \frac{c}{24}= \oint T_{zz} n^z, \quad \bar L_0 - \frac{c}{24} = \oint T_{\bar z \bar z} n^{\bar z}, 
\end{equation}
but in the limit as the circle becomes null, $n^{z} = 0$. Thus, for the CFT on the null circle, $L_0 - \frac{c}{24} = 0$, whether or not the right-moving component of the stress tensor vanishes. A finite right-moving energy density translates to a vanishing right-moving energy in the
limit because the proper size of the compact direction is going to zero. So for the near-horizon, near-extremal limit of the non-extreme black hole, we get the CFT on a null cylinder in a state with entanglement for both the left and right movers, but this still has $L_0 = \frac{c}{24}$. 

This entangled state is not a different candidate description of the self-dual orbifold; the state we have obtained in the near-horizon limit is in fact the same as the state we obtained above from the quotient perspective, just described in a different conformal frame on the boundary. The entanglement of the right-movers comes from the further coordinate transformation between $\tilde t$ and $v$ coordinates \eqref{bhctv}. In particular, the non-zero stress tensor $T_{\tilde t \tilde t}$ in \eqref{bhstress} can be seen to arise from applying the Schwarzian derivative formula \eqref{sder} to the conformal transformation between $\tilde t$ and $v$ coordinates. Thus, we obtain a consistent picture of the self-dual orbifold as dual to a particular entangled state in two copies of the CFT on the two boundaries. 

\subsection{Casual connection}
\label{causal}

We have obtained a description of the self-dual orbifold in terms of an entangled state in two copies of the CFT from two independent points of view. This description is also consistent with the description obtained in \cite{sd2} by considering the near-horizon limit of the extremal BTZ black hole. However, there is a problem with this description, as it fails to account for  the {\it causal} connections between the two boundaries, which would appear to imply direct interactions between the theories living on them.

To see that the boundaries at $z = \pm \infty$ are causally connected in the bulk, consider the metric \eqref{sdglobal}.  We see that the conserved quantity from $\phi$-translation invariance is $L = \dot \phi + \sinh 2 z \dot t$, and the minimum elapsed $t$ is along curves of $L=0$, for which along null geodesics
\begin{equation}
\Delta t = \int_{-\infty}^{\infty} \frac{dz}{\cosh 2z} = \frac{\pi}{2}. 
\end{equation}
If we think about \eqref{sdglobal} as a coordinate system on AdS$_3$,
there is no mystery about this causal connection: it corresponds to causal connections in the boundary. The part of the strip at $z = -\infty$ with $t > \pi/2$ is in the
causal future of the surface at $z = \infty$, $t=0$ in the boundary
geometry. The two parts of a surface of constant $t$ were offset in
$\tau - \theta$ by $\pi$, so when $\Delta t > \pi/2$, this offset is
overcome and the surfaces at $z = \pm \infty$ are connected by causal
curves in the boundary. From this AdS$_3$ point of view, this is not a
surprise; it is well-known that in AdS$_d$, points in the boundary
which are causally connected in the bulk are also causally connected
in the boundary: the bulk and boundary light cones agree for pure AdS
geometries. 

However, when we take the quotient, we must first delete from the
conformal boundary the points at $\tau + \theta = 0, \pi$, which are
fixed points of the identification acting on the conformal
boundary. This breaks the causal connection between the two strips on
the boundary at $z = \pm \infty$. This breaking of the explicit causal
connection does not immediately cause problems, as the connection
could be retained by a boundary condition linking the behaviour of
fields at $\tau + \theta = -\epsilon$ to the behaviour at $\tau +
\theta = + \epsilon$. However, when we make the
identification, we replace such a boundary condition with a periodic
boundary condition in the $\phi$ direction, and there is no longer any
connection between the behaviour of boundary fields on the two
strips. So in the quotient space, causal connection in the bulk is not reproduced by causal connection in the boundary.

This is a problem because an AdS/CFT 
calculation with causal connection in the bulk would usually predict a non-zero value for the commutator of operators on the two boundaries. If $\mathcal O_1$ is a scalar operator on
the boundary at $z =\infty$ and $\mathcal O_2$ is an insertion of the same scalar operator on the
boundary at $z = -\infty$,
\begin{equation} \label{corr}
\langle [\mathcal O_1(0), \mathcal O_2(\Delta t)] \rangle =
\Delta_{bulk}^\phi (\Delta t) \neq 0 \quad \mbox{for } \Delta t >
\pi/2,
\end{equation}
where $\Delta^\phi_{bulk}$ is the half advanced minus half retarded
propagator for the corresponding bulk field $\phi$. 

This non-trivial commutator between operators on the two boundaries
cannot arise simply from entanglement between the quantum states of
the theories on the two boundaries, as the expectation value of the
commutator is
independent of the state that the expectation value is evaluated
in. Thus, this seems to require some explicit interaction between the
two boundary theories. The bulk prediction \eqref{corr} is not consistent with our proposed description of the self-dual orbifold in terms of two independent, but entangled, boundary theories.

Indeed, it is very hard to see how the CFT description could be modified to produce such interactions. From the quotient perspective, we would expect local operators on the two strips to be simply independent once we delete the fixed points. From the near-horizon point of view it is even harder to see how some interactions could arise from taking
the DLCQ limit; the two CFTs should still simply be entangled. Specifically, in a black hole coordinate system, the portions of the boundaries that are captured are not in causal contact. So when we take this near-horizon limit, the commutator between fields on the two boundaries vanish in the region we are covering.

A possible resolution of our problem would be an obstruction to the extension of the CFT to the full boundary of the self-dual orbifold spacetime. However, there is a  barrier to finding such an obstruction. The bulk spacetime has an $SL(2,\mathbb R) \times U(1)$ isometry which acts transitively, mapping any point in the spacetime to any other point. Any obstruction to extending the geometry from the region covered by the black hole coordinates to the full spacetime must break this symmetry. The entangled state that we constructed preserves the spacetime isometries, as we would expect. We therefore expect the CFT in this state to live naturally on the conformal boundary of the full global self-dual orbifold spacetime, by the analogue of the argument of \cite{Luscher:1974ez} in the higher-dimensional case. Finite $SL(2,\mathbb R) \times U(1)$ actions can map a point on one boundary to any other point on that boundary in the full spacetime.

Another resolution would be a restriction on the types of correlation functions we can consider. The isometries can map a point on one boundary to any other point on that boundary, but will only map a pair of spacelike separated points to spacelike separated points. From the near-horizon point of view, we obtain correlation functions or commutators of operators on the two boundaries at spacelike separated points as a limit of the same observables in the theory on the boundary of the BTZ black hole. If we restrict to considering just such observables, there will be no conflict with our entanglement description even when we consider the full self-dual orbifold spacetime. 

Such a restriction may be necessary because of the instability of AdS$_2$ spacetimes observed in \cite{Maldacena:1998uz}. If we consider the back-reaction from adding some energy to the spacetime at one boundary, the spacetime will fail to be asymptotically AdS$_2$ to the future of the point where the energy is inserted. So we will not be able to impose asymptotically AdS$_2$ boundary conditions on the part of the other boundary causally connected to the point where the energy is inserted.  This suggests that we cannot consider correlation functions like \eqref{corr}, so the fact that a {\it linearised} bulk analysis predicts a value for this correlation function which is inconsistent with our entanglement description does not lead to actual inconsistencies in the full theory. This possibility thus seems plausible, but it would be valuable to understand the restrictions on which correlation functions we can consider in detail. It would be particularly useful to understand this from the CFT point of view. 

An alternative resolution is that there might not be any propagating states in the AdS$_2$ spacetime.  The authors of \cite{sd2} argued for such a picture by noting that the DLCQ field theory dual to AdS$_2$ does not have physical states charged under this $SL(2,\mathbb{R})$ group, so there should be no bulk states charged under the $SL(2,\mathbb{R})$ isometries of the spacetime.  In the absence of such propagating degrees of freedom, there can be no causal physical interaction between the two AdS$_2$ boundaries, again suggesting that the two boundary CFTs are non-interacting. From the CFT point of view, this would correspond to a claim that local operators like $\mathcal{O}_1$ do not create well-defined states in the chiral  CFT dual to the self-dual orbifold.

\section{Asymptotically Self-dual orbifold spacetimes}
\label{asymp}

We want to identify the self-dual orbifold geometry with a particular entangled state in a chiral CFT. The CFT should presumably have other states, and it is important to try to construct other geometries dual to these states. In this section, we discuss such constructions. We first discuss the boundary conditions defining what we mean by asymptotically self-dual orbifold. We then note that we can obtain another quotient geometry with a single boundary, which can be interpreted as the dual of the ground state of a single copy of the CFT. We then consider more general geometries constructed from the self-dual orbifold, first attempting a perturbative approach and then applying Garfinkle-Vachaspati solution-generating transformations. The more general solutions we construct have singularities in the bulk.

\subsection{Boundary conditions}
\label{bc}

Before looking for solutions, we must first specify the asymptotic boundary conditions we want to impose. We consider boundary conditions on a single conformal boundary, defining states in a single copy of the CFT. If we wanted to consider spacetimes dual to two copies of the CFT, they should have two conformal boundaries and satisfy these boundary conditions on each of them. 

Since the spacetime is locally AdS$_3$, we can impose the standard Brown-Henneaux boundary conditions \cite{Brown:1986nw}. These boundary conditions are
\begin{equation}
g_{uv} \sim e^{2r} + \mathcal{O}(1),  \quad g_{uu}, g_{vv}  \sim \mathcal{O}(1), \quad g_{ru}, g_{rv} \sim \mathcal{O}(e^{-2r}),\quad  g_{rr} \sim 1 + \mathcal{O}(e^{-2r}). 
\end{equation}
The leading term gives us the null cylinder metric on the conformal boundary if we assume the coordinate $u$ is periodically identified as in the self-dual orbifold solution. The subleading terms will then be interpreted as determining the stress tensor of the dual field theory. The subleading part of $g_{uv}$, which gives the trace of the stress tensor, vanishes when we satisfy the bulk equations of motion, so on-shell solutions actually have $g_{uv} \sim e^{2r} + \mathcal{O}(e^{-2r})$. 

However, in \cite{sd2}, a more restrictive boundary condition for asymptotically
self-dual orbifold spacetimes was proposed, requiring the $\mathcal{O}(1)$ part of $g_{vv}$ to vanish as well. This corresponds in the field theory to saying that the stress tensor component $T_{vv} = 0$. This was motivated by the chiral nature of the dual CFT. As reviewed in section \ref{review}, taking the near-horizon limit corresponds to a DLCQ limit in the field theory, which sets the right-movers to their vacuum state. If we interpret this as saying that the limiting theory dual to the self-dual orbifold has no right-moving excitations, it would be inconsistent to have a non-zero right-moving stress tensor. It is therefore appropriate to require that $T_{vv} = 0$ as part of the boundary conditions. We therefore propose that the dual description of a chiral CFT on the boundary is spacetimes with the boundary condition
\begin{equation} \label{cbc}
g_{uv} \sim e^{2r} + \mathcal{O}(1),  \quad g_{uu} \sim \mathcal{O}(1), \quad g_{ru}, g_{rv}, g_{vv}  \sim \mathcal{O}(e^{-2r}), \quad g_{rr} \sim 1 +\mathcal{O}(e^{-2r}). 
\end{equation}

Imposing the standard Brown-Henneaux boundary conditions would correspond to considering a non-chiral CFT on the null cylinder, where we retain some right-moving excitations. This is not the theory obtained in the strict near-horizon limit, but it remains interesting to consider it. It may be useful to consider situations where we do not take the strict near-horizon limit, and use the self-dual orbifold as an approximation to a region of the BTZ black hole spacetime \cite{Castro:2008ms,Castro:2010vi}. Since there are still some right-moving excitations, it may be that the Brown-Henneaux boundary conditions are then the appropriate ones to use to model the matching of the near-horizon region to the rest of the spacetime in this case. 

The choice of boundary conditions determines the asymptotic symmetries of the spacetime. For the standard Brown-Henneaux boundary conditions, the analysis of \cite{Brown:1986nw} tells us that the asymptotic symmetries are diffeomorphisms depending on two arbitrary functions $\xi^+(u)$, $\xi^-(v)$. The vector field generating the diffeomorphism is
\begin{align}
\xi^u &= 2 \xi^+(u) + \frac{1}{2} e^{-2r} \partial_v^2 \xi^-(v) + \mathcal{O}(e^{-4r}), \\
\xi^v &= 2 \xi^-(v) + \frac{1}{2} e^{-2r} \partial_u^2 \xi^+(u) + \mathcal{O}(e^{-4r}), \\
\xi^r &= - \partial_u \xi^+(u) - \partial_v \xi^-(v) + \mathcal{O}(e^{-2r}). 
\end{align}
Since $u$ is a compact coordinate, $\xi^+(u)$ is a periodic function, and can be expanded in terms of modes which satisfy a Virasoro algebra. This includes the left-moving $U(1)$ symmetry of the self-dual orbifold given by $\partial_u$. However, as $v$ is non-compact, $\xi^-(v)$ is not periodic, and the right-moving symmetry here is not simply a Virasoro algebra. Its interpretation from the CFT point of view is somewhat unclear.

For the boundary conditions \eqref{cbc}, the asymptotic isometries are restricted. As shown in \cite{sd2}, only the diffeomorphisms with $\xi^-(v) = A + B v + C v^2$ survive. These correspond to the $SL(2,\mathbb{R})$ Killing vectors \eqref{PKV}. Thus, for the boundary conditions \eqref{cbc}, the asymptotic isometries are $SL(2,\mathbb{R}) \times$ Virasoro, where the Virasoro contains the left-moving $U(1)$ symmetry. 

Note that the diffeomorphisms which transform between the $u,v,r$ coordinates we are using and the orbifold global coordinates $t,\phi,z$ or black hole coordinates $\tilde t, \tilde \phi, \bar r$ do not satisfy the boundary condition \eqref{cbc}.  Rather the boundary conditions are also transformed by these diffeomorphisms and must be expressed in the new coordinate frame.  From the boundary point of view, the coordinate transformations to global and black hole corrdinates corresponded to conformal rescalings of the $v$ coordinate. But with the boundary conditions \eqref{cbc}, the CFT only has a conformal symmetry acting on the $u$ coordinate. The conformal transformations of the $v$ coordinate are a part of the symmetries in $\xi^-(v)$ which is not in the $SL(2,\mathbb{R})$ subgroup we retain.  Thus, while we are free to make such conformal transformations, the theory will not be invariant under the change of variables.   In view of this, when we look for asymptotically self-dual orbifold solutions satisfying the boundary conditions \eqref{cbc}, we will  study them in the analogue of the $(u,v,r)$ coordinates only.

\subsection{Ground state of a chiral CFT and the null orbifold}
\label{single}

In \cite{FigueroaOFarrill:2004bz}, a general classification of causally well-behaved quotients of AdS$_3$ was given. There was one other quotient which had the
same type of boundary metric as the self-dual orbifold, namely the quotient  by 
\begin{equation} \label{gsorb}
\xi = (V-Y)(\partial_U + \partial_X) - (U-X)(\partial_V + \partial_Y), 
\end{equation}
where $U,V,X,Y$ are the coordinates on the $\mathbb R^{2,2}$
embedding space. This Killing vector has $||\xi||^2 = 0$, but the Killing vector never vanishes, so the quotient has no fixed points in the spacetime. However, when we consider the action just on AdS$_3$, the resulting quotient space will contain closed null curves. This Killing vector lies in one of the two $SL(2,\mathbb{R})$ factors in the  $SL(2,\mathbb{R}) \times SL(2,\mathbb{R})$ isometry group, so the quotient has an $SL(2,\mathbb{R}) \times U(1)$ isometry, as for the self-dual orbifold. A coordinate system which covers the whole spacetime  is \cite{FigueroaOFarrill:2004bz}
\begin{align} 
U+X &= e^{-\rho} \sin v + 2 u e^\rho \cos v, \\
U-X &= e^\rho \sin v, \\
V+Y &= e^{-\rho} \cos v - 2 u e^\rho \sin v, \\
V-Y &= e^\rho \cos v. 
\end{align}
In these coordinates, $\xi = \partial_u$ and the metric takes the form 
\begin{equation} \label{gsmet}
ds^2 = -dv^2 + d\rho^2 - 2 e^{2\rho} du dv.
\end{equation}
The quotient in these coordinates is an identification $u \sim u + 2\pi$. Since the quotient has no fixed points, the resulting quotient spacetime is smooth. The spacetime has a single  boundary at $\rho \to \infty$. The metric on this boundary is a null cylinder, as in the self-dual orbifold. In terms of the global coordinates on the boundary of AdS$_3$, the quotient has a single line of fixed points at $\tau + \theta = \pi$. The supersymmetry of this solution was analysed in \cite{FigueroaOFarrill:2004yd}, where it was shown that taking this
quotient of AdS preserves 3/4 of the SUSY, including 1/2 of the
left-moving SUSY. Note that although the geometry has an $SL(2,\mathbb{R}) \times U(1)$ isometry, it does not have an AdS$_2$ factor.

This quotient can also be viewed as an
identification along a null direction in Poincar\'e coordinates. That is, if we introduce the standard Poincar\'e coordinates $x^+, x^-, Z$ on AdS$_3$, in terms of which the metric is 
\begin{equation} \label{gsmet2}
ds^2 = \frac{-2 dx^+ dx^- + dZ^2}{Z^2}, 
\end{equation}
then $\xi = \partial_{x^+}$. This Poincar\'e coordinate system only covers a region of the spacetime, but it has the advantage that the spacetime written in these coordinates satisfies the more restrictive boundary conditions of section \ref{bc}.  This solution and the self-dual orbifold are the only locally AdS$_3$ spacetimes whose boundary metrics are null cylinders. 

This geometry can also be obtained by taking the near-horizon limit of the
$M=0$ BTZ black hole: if we start with the Poincar\'e coordinates \eqref{gsmet2}, 
the $M=0$ BTZ black hole is obtained by making the identifications
$(x^+, x^-) \sim (x^+ + 2\pi, x^- - 2\pi)$. We take the
near-horizon limit by defining $x^+ = \tilde x^+$, $x^- =
e^{-2\rho_0} \tilde x^-$, $Z = e^{\rho_0} \tilde Z  $, and
take $\rho_0 \to -\infty$ for fixed $\tilde x^+, \tilde x^-, \tilde
Z$. This gives a metric of the same form, but with $\tilde x^+ \sim \tilde x^+ + 2\pi$, giving the null quotient (\ref{gsmet2}). 

Since the geometry has a single boundary, we would interpret this spacetime as the dual description of a single copy of the CFT living on the null cylinder in some pure state. We can identify the appropriate state by proceeding as in the self-dual orbifold, taking the $SL(2,\mathbb{R}) \times SL(2,\mathbb{R})$ invariant vacuum state on the boundary of AdS$_3$ and considering the quotient action on it. The appropriate coordinate system in this case is just the Poincar\'e coordinates \eqref{gsmet2}. We know that the state dual to AdS$_3$ in Poincar\'e coordinates is a ground state for both the left- and right-moving modes, so we propose that the dual of this quotient spacetime is the same ground state with the momentum for left-moving excitations quantised, breaking the symmetry of the state to $SL(2,\mathbb{R}) \times U(1)$. This is consistent with obtaining the quotient as the near-horizon limit of the $M=0$ BTZ black hole; as the black hole mass goes to zero, the temperature for both left- and right-moving modes vanishes. Thus, the dual CFT interpretation of \eqref{gsmet} is as a saddle-point for a single copy of the CFT on the null cylinder in a ground state.  (See the related discussion in \cite{joanetal}.)

The main problem with this discussion is that the spacetime contains
closed null curves (CNCs), so one might not expect the spacetime
\eqref{gsmet} to be a good description of the boundary field theory
state. In particular, winding string modes wrapping this compact direction will be light and could produce important corrections to the geometry. The unbroken supersymmetry in this spacetime may protect the geometry from such corrections, however. This problem is highly reminiscent of the Schr\"odinger
spacetimes \cite{Son:2008ye,Balasubramanian:2008dm}, which similarly contain closed null curves (as remarked in \cite{Maldacena:2008wh}). The self-dual
orbifold would then be thought of as analogous to the finite-temperature versions of
Schr\"odinger spacetimes \cite{Maldacena:2008wh,Adams:2008wt,Herzog:2008wg}, in that the circle becomes spacelike
everywhere in the bulk. The situation is slightly actually slightly better than in the Schr\"odinger case, as the circle becomes constant size in the bulk, whereas it was asymptotically null in the finite-temperature Schr\"odinger spacetimes.       It is interesting that while the metric (\ref{gsmet}) has CNCs, adding a tiny temperature on the left or right moving side of the dual field theory seems to regulate the CNCs.    A purely left or right-moving temperature corresponds to extremal rotation in the BTZ black hole whose near-horizon limit we are examining.   This is reminiscent of the ``desingularization by rotation" in \cite{Balasubramanian:2000rt,Maldacena:2000dr}.

In \cite{FigueroaOFarrill:2004bz}, the problem
with the causal structure of (\ref{gsmet}) was formally resolved by combining the quotient action on AdS$_3$ with an action on the $S^3$ to obtain an action which was everywhere
spacelike. This removes the closed null curves, but the resulting
spacetime is not stably causal, as a
$\mathbb{Z}$ identification on a compact space like $S^3$ will identify pairs
of points which are arbitrarily close together. As a result, even if
we include such an action on the $S^3$, we still need to worry about
light states associated with strings winding around the circle: there
will be winding sectors where these strings are arbitrarily light. 

Nonetheless, as in the Schr\"odinger case, this quotient spacetime provides an interesting simple example of the dual of a pure state, and it is worth considering as at least a formal dual of the ground state. The quotients with actions on the sphere are also interesting, as they
should correspond to ensembles where in addition to the temperature we
are turning on a chemical potential for some $R$-charge.

\subsection{Perturbative excitations}

From the CFT point of view, we would expect to be able to consider
arbitrary states for the left-movers. These might not all have a
geometrical interpretation, but small excitations around the state
corresponding to the self-dual orbifold might be expected to
correspond to perturbations around the self-dual orbifold geometry. We are most interested in understanding chiral excitations, which will
satisfy the boundary conditions of \cite{sd2}, and account for the
entropy of the original black hole. In this section, we consider such chiral excitations on the full extremal BTZ black hole geometry. We find that surprisingly, excitations which keep the right-movers in their ground states cannot be consistently described by  small perturbations around the black hole spacetime.\footnote{Since the self-dual orbifold is a near-horizon limit of BTZ, this immediately implies that such chiral excitations also cannot be described perturbatively on the self-dual orbifold.}

We consider linearised fields on the extremal BTZ black hole background. We will start by considering scalar fields. We consider a scalar field $\Phi$ of mass $\mu^2$,
and write the field in Fourier modes as
\begin{equation}
\Phi = e^{i \omega t + i m \phi} f_{\omega m}(r),
\end{equation}
where we are working in the BTZ black hole coordinates defined in
\eqref{btz}. For the extremal BTZ black hole, $r_+ = r_-$,
chiral excitations from the CFT point of view correspond
to considering co-rotating modes with $\omega = -m$ in the bulk spacetime. 

It turns out that precisely these modes are not regular on the
horizon. The field equation is $\Box \Phi - \mu^2 \Phi=0$, and on the
BTZ black hole \eqref{btz}, if $\Phi = e^{i \omega t + i m \phi}
f(r)$, 
\begin{equation}
\Box \Phi = \frac{1}{r^2 h} [ r^2 \omega^2 - (r^2 -r_+^2 - r_-^2) m^2
+ 2 r_+ r_- \omega m]  e^{i \omega t + i m \phi}
f(r) +   e^{i \omega t + i m \phi} \frac{1}{r} \partial_r ( r
h \partial_r f(r)), 
\end{equation}
where $h(r) = (r^2-r_+^2)(r^2-r_-)^2/r^2$. Hence, if $r_+ = r_-$, and
$\omega = -m$, the first term disappears and $\Box \Phi = e^{i \omega
  t + i m \phi} \frac{1}{r} \partial_r ( r h \partial_r f(r))$.  The
solution of the radial equation which satisfies the boundary
conditions at infinity in this case is then just 
\begin{equation} \label{scalar2}
f_{\omega = -m}(r) = c_m (r^2 - r_+^2)^{-h_+/2}, 
\end{equation}
where $h_+ = \frac{1}{2} (1 + \sqrt{1 + \mu^2)})$. The surprising
feature of this solution is that it blows up as we approach the black
hole horizon at $r \to r_+$. This indicates that if we want to
consider chiral modes on the BTZ boundary,
this perturbative analysis will break down. 

In fact, this failure is analogous to the ``no-hair'' theorem for the
non-rotating black hole, which says that there is no regular solution
for the scalar field with $\omega =0$. The connection can be seen more
clearly by considering the general BTZ black hole, with $r_+ \neq
r_-$. Then if we define $\omega_c = \omega + \frac{r_-}{r_+} m$, 
\begin{equation}
r^2 \omega^2 - (r^2 -r_+^2 - r_-^2) m^2
+ 2 r_+ r_- \omega m = r^2 \omega_c^2 - 2 \frac{r_-}{r_+} (r^2- r_+^2)
\omega_c m - \frac{r_+^2-r_-^2}{r_+^2} (r^2-r_+^2) m^2. 
\end{equation}
So if $\omega_c=0$, this factor vanishes on the horizon. The radial
equation can be rewritten in terms of an effective potential by
introducing a tortoise coordinate $r_*$ such that $dr_* = h^{-1} dr$;
then writing $f(r) = r^{-1/2} \psi(r)$, the radial equation becomes 
\begin{equation}
\partial_{r_*}^2 \psi + \omega_c^2 \psi - h(r) v_{eff}(r) \psi = 0, 
\end{equation}
where $v_{eff}(r) >0$ for all $r$. Because of the overall $h(r)$
factor, the effective potential contribution vanishes near the
horizon, so for $\omega_c \neq 0$, the solutions near the horizon will
look like $e^{\pm i \omega_c r_*}$, giving the usual ingoing and
outgoing modes on the horizon. But if $\omega_c = 0$, the solutions
will grow or decay near the horizon, as $v_{eff}(r) >0$. The solution
which is regular at infinity will (at least generically) include a
growing part near the horizon. 

For $r_- = 0$, this is the usual argument that there are no static
scalar hairs on the black hole. A similar interpretation in our case
would be that the black hole cannot support a chiral
perturbation of the scalar field.

This calculation can trivially be extended to the vector case by
observing that since we are in $2+1$ dimensions, a vector field is dual
to a scalar. Thus, if we want a solution for a Maxwell field with
field strength given by $F$, we can find it by writing $F = \star d
\Phi$ for a scalar $\Phi$ satisfying the massless wave equation. The
scalar solution of \eqref{scalar2} gives a field strength which blows
up on the horizon.

\subsection{Non-thermal states of chiral CFT and traveling wave solutions}
\label{wave}

Having failed to construct more general geometries perturbatively, we will now consider applying a solution-generating transformation to obtain new solutions of the full equations of motion.
Both the extremal BTZ black hole and the self-dual orbifold have a
null Killing vector field, given by $\partial_v$ in the $(u,v,r)$
coordinates. We can therefore apply the Garfinkle-Vachaspati solution
generating transformation \cite{Garfinkle:1990jq} to add a travelling
wave, as was done for asymptotically flat black string solutions in
\cite{Kaloper:1996hr}. The null Killing vector is $k = \partial_v$, so
with the index lowered $k = e^{2r} du$. This satisfies $\nabla_{[\mu}
k_{\nu]} = k_{[\mu} \nabla_{\nu]} S$ with $S = -2r$. The Garfinkle-Vachaspati
technique tells us that we can generate a new solution $\tilde
g_{\mu\nu}$ by choosing a function $\Psi$ satisfying $\partial_v \Psi
=0$ and $\nabla^2 \Psi = 0$, and defining 
\begin{equation}
\tilde g_{\mu\nu} = g_{\mu\nu} + e^S \Psi k_\mu k_\nu.
\end{equation}
That is, 
\begin{equation}
\tilde{ds}^2 = (1 + e^{2r} \Psi) du^2 + 2e^{2r} du dv + dr^2.
\end{equation}
This spacetime will be asymptotically AdS$_3$ or asymptotically
self-dual orbifold depending on whether we make the spacelike
direction $u+v$ or the null direction $u$ compact. 

If we consider just the three-dimensional spacetime, then $\nabla^2 \Psi =
e^{-2r} \partial_r(e^{2r} \partial_r \Psi)$, and $\Psi = f_0(u) +
  f_1(u) e^{-2r}$. To preserve the asymptotics of the spacetime, we
  should set $f_0(u)= 0$; the solution is then 
\begin{equation}
\tilde{ds}^2 = (1 + f_1(u)) du^2 + 2e^{2r} du dv + dr^2.
\end{equation}
However, this transformation is trivial; the Garfinkle-Vachaspati
transformation in general adds a gravitational wave to the previous
solution, but in three dimensions, there is no gravitational
radiation. That is, any solution of the vacuum equations of motion in $2+1$ dimensions is locally AdS$_3$. Thus, this solution is just a locally AdS$_3$ spacetime
written in an unfamiliar coordinate system.\footnote{Note that the
  situation here is different from the full asymptotically flat
  solution considered in \cite{Kaloper:1996hr}, where adding the $l=0$
  mode produces a physically different solution. This implies that
  these different solutions all have the same near-horizon BTZ
  region.}

To obtain something non-trivial we need to introduce some additional directions and allow $\Psi$ to depend them. Consider for example taking the product of our geometry with an $S^3$, as in the simplest embeddings in string theory, and allowing $\Psi$ to depend on
the coordinates on the $S^3$ factor in the geometry. For simplicity, assume $\Psi$ is in a
particular spherical harmonic on the sphere, so $\Psi = Y_{lm}(\theta,
\phi,\psi) g(r,u)$. Then 
\begin{equation}
e^{-2r} \partial_r (e^{2r} \partial_r g) - l(l+2) g = 0,
\end{equation}
with solutions $g(r,u) = f_0(u) e^{lr} + f_1(u) e^{-(l+2) r}$. As
previously, take $f_0(u) = 0$ to preserve the boundary conditions, and
for each spherical harmonic we have one functions worth of
solutions. For example, if we take the harmonic with $l=2, m=0$, we
have a solution
\begin{equation}
\tilde{ds}^2 = (1 + f_1(u) \cos 2\theta e^{-2r}) du^2 + 2e^{2r} du dv + dr^2 +
d\theta^2 + \sin^2 \theta d\phi^2 + \cos^2 \theta d\psi^2.
\end{equation}
This solution has a non-vanishing Weyl tensor, indicating the presence
of a gravitational wave, and showing explicitly that this is a
non-trivial example of an asymptotically self-dual orbifold spacetime. As in
\cite{Kaloper:1996hr}, these geometries are singular: they have diverging Riemann tensor components on the would-be
horizon at $r \to - \infty$, although the curvature invariants are finite.  Thus they have no
extension to global coordinates and only satisfy the self-dual
orbifold boundary conditions on one boundary, in the region
corresponding to the $(u,v,r)$ coordinates. The deformed spacetime breaks the $SL(2,\mathbb{R})$ symmetry, and does not have an AdS$_2$ factor,
although if we made a dimensional reduction to two dimensions, the
geometry would be asymptotically AdS$_2$ in Poincar\' e coordinates.

It is very satisfying that we finally have some examples of
asymptotically self-dual orbifold spacetimes, even if they have mild singularities in the bulk. It would be interesting to understand their dual description. We expect them to correspond to more general states for the left-movers, where the operators dual to excitations on the sphere have
non-zero expectation values, and a ground state for the right-movers.  However, the travelling wave breaks the symmetry corresponding to $L_{-1}$ in $SL(2,\mathbb{R})$. We think of this symmetry as acting on the right-movers, so these geometries do not correspond to not precisely the same right-moving ground state as in the self-dual orbifold. (Since the geometry is still invariant under $L_0$ and $L_{1}$, it still corresponds to a ground state. This is also clear from the fact that it satisfies the boundary conditions of \cite{sd2}.)
It's surprising that the Garfinkle-Vachaspati transformation breaks this symmetry; we would naively have thought of it as acting just on the left-movers. It would be interesting to understand this in more detail. 

\subsubsection{Other AdS$_2$ cases}

This solution-generating transformation provides a useful way to generate new solutions. It is therefore interesting to ask if it is special to the
self-dual orbifold, or can be applied in other contexts where the
geometry has an AdS$_2$ factor. The NHEK geometry does not have a null
Killing vector, so it cannot be applied in that case (the analogue of
the Killing vector considered here would be $\partial_t$, which is not
everywhere null because of the fibration over $\theta$). Thus, there are no such solutions in NHEK, as we might have expected given the results of \cite{Amsel:2009ev,Dias:2009ex}. 

In the
context of Reissner-Nordstr\"om AdS black holes, the near-horizon
geometry in for example AdS$_4$ is 
\begin{equation}
ds^2 = \frac{l_2^2}{\rho^2} (-dt^2 + d\rho^2) + d\vec{x}^2,
\end{equation}
where $l_2 = l/\sqrt{6}$, with a vector field $A =
\frac{g_F}{\sqrt{12} \rho} dt$. Clearly there is no null Killing
vector here, but one might hope to find one in the uplift to the full string or M theory geometry. In the self-dual orbifold case, the $v$ direction is timelike in the AdS$_2$ factor, but becomes null when we uplift it to the three-dimensional geometry.

Consider for definiteness the uplift to eleven dimensions given in \cite{Gauntlett:2009bh}. If we work in units where the
$S^7$ has unit radius, $l = 1/2$, $g_F = 1/2$, and the
eleven-dimensional metric is 
\begin{equation}
ds^2_{11} = ds^2_4 + ds^2_{CP^3} + (\eta + A)^2, 
\end{equation}
where $\eta$ is the one-form dual to the Reeb vector in the writing of
$S^7$ as a Hopf fibration over $\mathbb{C}P^3$. Hence in
eleven dimensions there is a partial
cancellation between the two factors as in the self-dual orbifold case, but
\begin{equation}
  g_{tt} = -\frac{l_2^2}{\rho^2} + \frac{g_F^2}{12 \rho^2}  = -
  \frac{1}{48 \rho^2},
\end{equation}
so the Killing vector $\partial_t$ remains timelike, and we can't
apply the Garfinkle-Vachaspati transformation to this solution. The situation for
Reissner-Nordstr\"om AdS$_5$ black holes is the same; it seems to be only when
we are uplifting from AdS$_2$ to a three-dimensional solution that the
factors work out so that we get a null isometry in the
higher-dimensional geometry.

This suggests that there is something a little special about the
AdS$_2$ from the dual CFT point of view in the self-dual orbifold
case; the null structure that is responsible for the chiral CFT
interpretation here isn't obviously present in higher-dimensional
cases.

\section{Discussion}
\label{concl}

We have studied the description of the self-dual orbifold from the point of view of the dual CFT, and constructed examples of asymptotically self-dual orbifold spacetimes, which should be dual to other states of the CFT. We have proposed that the full spacetime can be described as an entangled state in two copies of the CFT, living on the two boundaries. This description appears to have problems with the causal connection between the two boundaries, which would lead to predictions for bulk correlation functions which cannot be reproduced by considering an entangled state. However, the special nature of AdS$_2$ suggests that there will be restrictions on the correlation functions which can be consistently considered. Acting with an operator in the field theory to throw in some energy from the boundary will cause a back-reaction  which violates the boundary conditions on the boundary after the  operator insertion. We have suggested that the problematic correlations involving timelike separated operators on the two boundaries may simply not be legitimate observables. This issue needs further exploration. 

We discussed the asymptotic boundary conditions for the spacetime. Following \cite{sd2}, we argued that considering a chiral CFT on the boundary is associated with a boundary condition that is more restrictive that the ones imposed by Brown and Henneaux. We constructed examples of asymptotically self-dual orbifold spacetimes satisfying this boundary condition. This is interesting as it can be challenging to construct asymptotically AdS$_2$ spacetimes; in the Kerr-CFT context it was shown in \cite{Amsel:2009ev,Dias:2009ex} that the only spacetimes satisfying the relevant boundary condition are diffeomorphic to the background. Note however that the solutions we construct are singular in the bulk. Other examples of asymptotically AdS$_2$ spacetimes which are regular were recently obtained in higher-dimensional contexts in \cite{Faulkner:2010rt} by considering RG flows from one AdS$_2$ to another.    We also identified a geometry corresponding to the ground state of the chiral CFT which is dual at finite temperature to the self-dual orbifold.  This new geometry is obtained as a near-horizon limit of the $M=0$ BTZ black hole, just as the self-dual orbifold is the near-horizon limit of the $M > 0$  extremal BTZ black holes. 

In the simplest embedding in string theory, we would consider the self-dual orbifold geometry $\times S^3 \times T^4$. As discussed in \cite{FigueroaOFarrill:2004bz}, in this context we can generalise the quotients considered here by adding an action on the $S^3$. This corresponds to introducing a chemical potential for an $R$-charge. For the orbifold of section \ref{single}, corresponding to the CFT in a ground state for both the left- and right-moving excitations, it is meaningful to introduce such a chemical potential because we are considering the theory in a Ramond sector, so there is a degenerate set of ground states, which carry different $R$-charge. 

We have assumed throughout this work that we were considering the field theory in the Ramond sector, so there is some unbroken supersymmetry in the solution. Since the circle is of finite size everywhere in the spacetime, we can change our choice of spin structure, which corresponds to considering the field theory in the Neveu-Schwarz sector. If the compact circle in the interior is smaller than the string scale, this solution will then have a tachyon. Since this circle has the same size everywhere, we would expect the condensation of this tachyon to destroy the whole spacetime. This tachyon condensation process for the full extreme BTZ geometry was considered in \cite{Parsons:2009si}.
 
\section*{Acknowledgements}

We are grateful for useful discussions with Jan de Boer, Alejandra Castro, Veronika Hubeny, Alex Maloney, Mukund Rangamani, Joan Simon and especially Don Marolf, who suggested the calculation in section \ref{wave}. JP and SFR are partially supported by the STFC. SFR thanks the Galileo Galilei Institute for hospitality and the INFN for partial support during the completion of this work.  VB is supported by DOE grant DE-FG02-95ER20893 and was partially supported by NSF PHY05-51164 to the KITP.  VB thanks  the organizers of the Amsterdam Summer Workshop in String Theory, as well as the Aspen Center for Physics, the Santa Fe Institute, and the KITP in Santa Barbara for hospitality while  this work was being carried out.

\bibliographystyle{utphys}
\bibliography{sdorbifold-2}

\end{document}